# Exploration of the deep geothermal potential of Petite-Terre Island in Mayotte

**C. Dezayes, A. Stopin, P. Wawrzyniak, F. Gal, T. Farlotti, P. Calcagno, A. Armandine les Landes**

**BRGM, Georesources Division, 3 avenue C. Guillemin, 45060 Orléans, France**

*c.dezayes@brgm.fr*

## ABSTRACT

In 2017, the electricity mix of the Department of Mayotte included 5% photovoltaic production, spread over more than 70 installations on the island, and 95% diesel thermal production, provided by two power plants. Nevertheless, since 2005, in order to diversify its electricity supply, the local authorities of Mayotte set up a vast program to develop renewable energies, including the evaluation of Mayotte's geothermal potential to produce electricity.

In this framework, BRGM started a program to explore the deep geothermal potential of the department of Mayotte. The conclusions of the different studies confirmed the idea of a strong potential under Petite Terre (Traineau *et al.*, 2006; Sanjuan *et al.*, 2008; Darnet *et al.*, 2019).

Our present study aims at refining the knowledge of the subsurface and the geothermal system potentially present beyond a depth of 1000m. This new phase, carried out between 2021 and 2022 includes a geophysical measurement campaign accompanied by a fracture analysis and the study of gas emanating from the ground.

The different data set acquired during the campaigns are analysed and interpreted jointly with the data acquired in the framework of the monitoring of the exceptional seismo-volcanic activity observed since May 2018 off the coast of Mayotte.

The result of this analysis is a 3D geological model of the structure of the island of Petite Terre, completed by a hydrothermal model of the functioning of the geothermal system, with the aim to help in locating the most favourable zones for geothermal exploitation.

## 1. Introduction

In 2017, the electricity mix of the Department of Mayotte included 5% photovoltaic production, spread over more than 70 installations on the island, and 95% diesel thermal production, provided by two power plants. Nevertheless, since 2005, in order to diversify its electricity supply, the local authorities of Mayotte set up a vast program to develop renewable energies, including the evaluation of Mayotte's geothermal potential to produce electricity.

In this framework, BRGM started a program to explore the deep geothermal potential of the department of Mayotte. The conclusions of the different studies confirmed the idea of a strong potential under Petite Terre (Traineau *et al.*, 2006; Sanjuan *et al.*, 2008; Darnet *et al.*, 2019).

Our present study aims at refining the knowledge of the subsurface and the geothermal system potentially presents beyond a depth of 1000m. This new phase, carried out between 2021 and 2022 including field campaigns accompanied by a 3D geological model multi-data gathering and hydrological simulation in order to define the best place to the first exploration drilling.

This paper shortly presents results of field campaigns and a 3D geological model, the other phase of the study being on going.

## 2. Geological context

Mayotte is a volcanic island lying on the ocean floor of the southern Somali Basin between Africa and Madagascar belonging to the Archipelago (Figure 1). It is mainly composed of volcanic formations and surrounded by the largest closed lagoon in the Indian Ocean, bounded by a reef barrier. The main lithologies observed at the surface are stacked basalt, nepheline and tephrite flows, phonolite domes and pyroclastic deposits dating less than 5Ma (Figure 2; Nehlig *et al.*, 2013; Lacquement *et al.*, 2013).

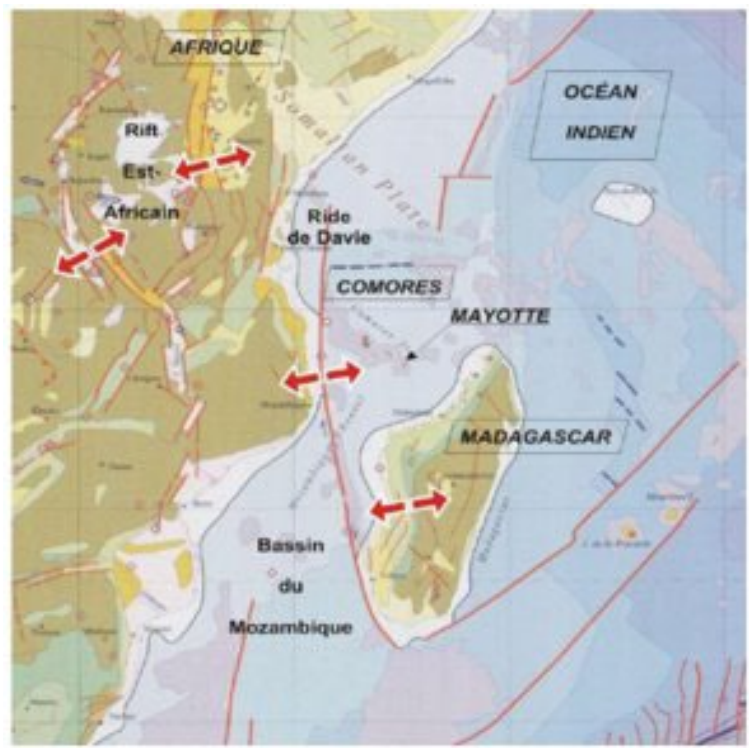

**Figure 1 : Sismotectonic map from CCGM-UNESCO, 2002**

## 3. Methodology

Investigations of Mayotte island geothermal potential has started in 2000's by geological and geochemical studies (Traineau *et al.*, 2006), then geophysical surveys (Pajot *et al.*, 2007). This latter study highlights a heavy and magnetic body under Petite Terre Island close to the Airport Beach, where $CO_2$ gas emanations occur. These news results allow to complete and confirm the previous ones (Sanjuan *et al.*, 2008). Based on these results, a more detailed exploration program has been defined (Darnet *et al.*, 2019) and new investigations have been done during the summer 2021.

These complementary surveys focused on Petite Terre Island in order to acquire:

- geological data, and in particular a detailed analysis of the fracturing and the mineralogy associated with this fracturing in order to estimate the permeability of the rocks and to identify the zones most likely to be permeable;

- geochemical data, with a study of the geothermometry of the gases in order to better estimate the temperature of the source of these gases and the associated hydrothermal system;

- additionnal magnetotelluric data onshore and offshore to image in three dimensions the electrical conductivity of the subsurface and to specify the geometry of the hydrothermal system.

All data, previous and recent ones, has been included together in GeoModeller™, a 3D geological modeller developed by BRGM. This software allows a coherent interpretation in three dimensions of geological objects, based on different types of data to constrain their geometries. We obtain a 3D representation of geological formations, which will serve as a basis for hydrothermal simulations using the ComPASS platform, also developed by BRGM (https://charms.gitlabpages.inria.fr/ComPASS).

## 4. Results

The main results are summarized below.

### 4.1 Geological data

The structural analysis of fractures observed in various location of Mayotte shows four main sets: N140°E ± 20°; N90°E ± 20°; N0°E ± 20°; N50°E ± 20° (Figure 2).

The N140°E ± 20° (or NW-SE) direction is the major direction and is in relation to the regional geodynamic interpretation, which consider the Comoros archipelago as a large Riedel shear structure (Famin *et al.*, 2020).

The fracture density is around 1fr/m and the distribution is random with few clustering. This can indicate a low formation permeability with a probable anisotropy.

No filling elements have been found in the fracture, except rare alteration.

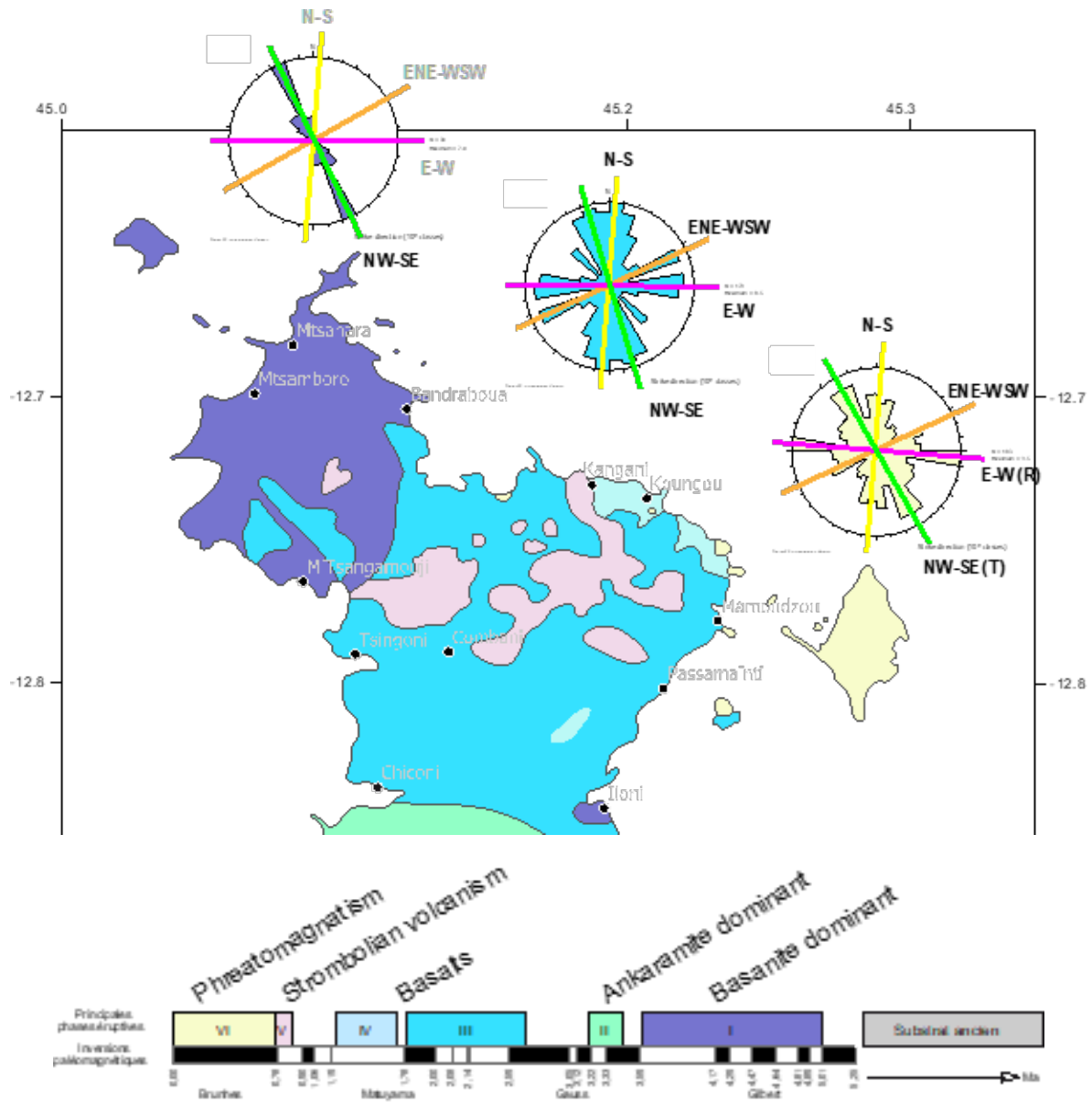

**Figure 2: Fracture directions found in the different volcanic formations and main sets highlighted (background : volcano-structural map of Mayotte, Nehlig et al. (2013)).**

## 4.2 Geochemical analysis

The main magmatic emanation is present on the Airport Beach of Petite Terre Island. Gas analysis shows that $CO_2$ is deep and of magmatic origin.

Geothermometer methods have been applied by various teams (Sanjuan *et al.*, 2008; Liuzzo *et al.*, 2021; this study; Figure 3) and assess an equilibrium temperature higher than 250°C, which could be the geothermal reservoir temperature at depth.

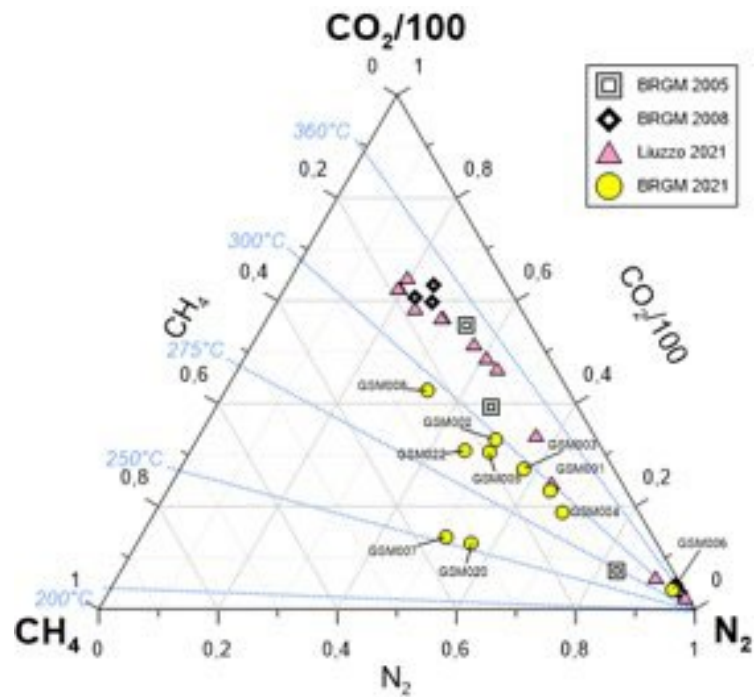

**Figure 3: CO2-N2-CH4 ternary diagram (Giggenbach *et al.*, 1991) with recent and previous measurements plotted. Temperature lines reported in blue are the equilibrium temperatures given by the CH4/CO2 geothermometer.**

We also measured the gas flux directly on various location and visually estimated the gas flux by walking on the beach. Based on that, we drawn a $CO_2$ flux map that shows a NW-SE trend equivalent to that of the fractures. This could highlight a large structure with NW-SE direction that allow the magmatic fluid upwelling. This structure have the same direction as the main fracture set and then could constitute a major driving element for the geothermal system.

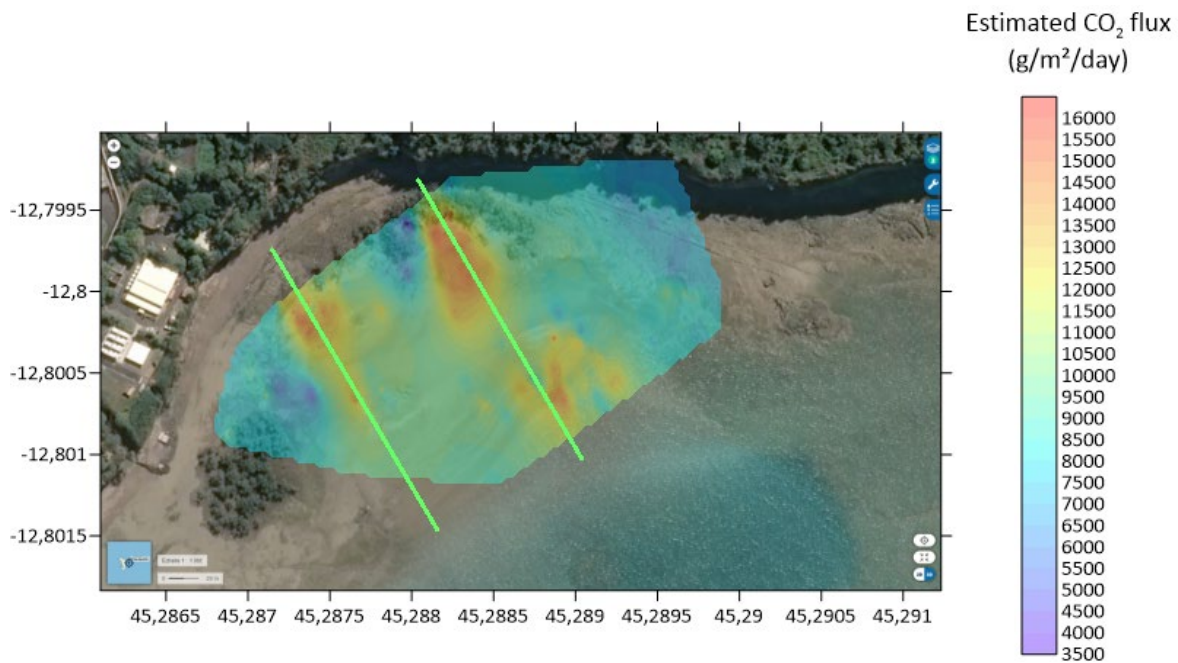

**Figure 4: Map of gas flux assessment on the airport beach. Green lines: NW-SE trend tendency**

## 4.3 Geophysical survey

Inversion of 32 land (Metronix ADU MT stations) + marine MT (Mappem Geophysics Statem stations) soundings with MININ3D code (Hautot et al., 2000, Hautot et al., 2007) provide a 3D resistivity model (Figure 5). These images display a conductive structure overlaying a resistive body with a varying depth interface (600m below Moya beach and 1.6km below the airport (Figure 5). Similarities with the Johnston *et al.* (1992) model can be established but with a more complex geometry. We can interpret this interface as the frontier between a geothermal reservoir and its caprock. On the map view at 2432m depth, the resistivity body has a NW-SE global trend (Figure 5). This overall good quality of the hybrid land + sea MT measurement permitted to give a well-constrained 3D imaging of a potential geothermal target.

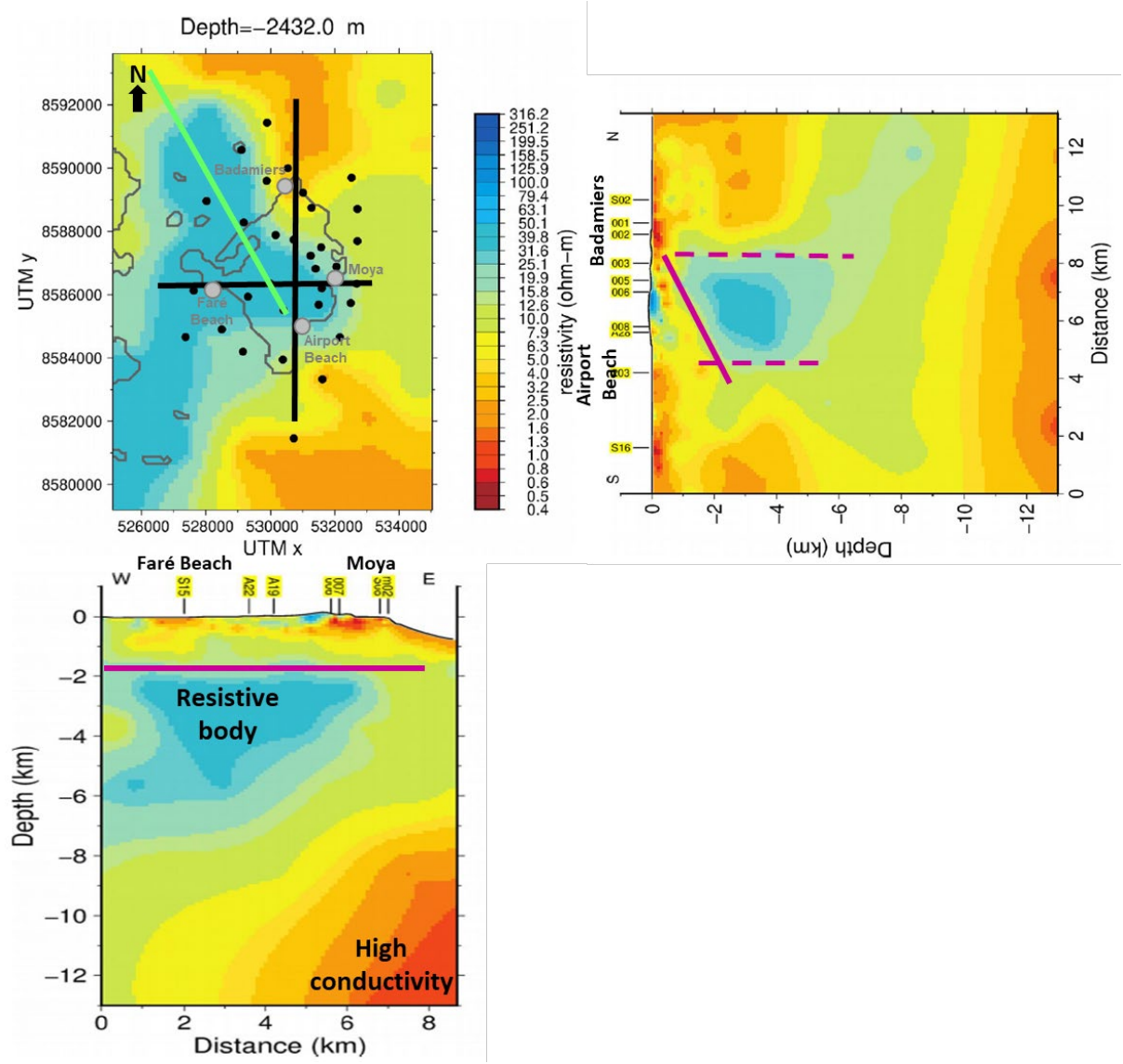

**Figure 5: Resistivity map and cross-section on Petite Terre Island. On the map, black points: station location, black lines: location of cross-section, green line: SW-SE trend tendency. On cross-section, purple lines: limitation of resistive body.**

## 4.4 3D geological model and hydrothermal conceptual model

Resistivity inversion has been used as base of the 3D geological model in order to define the main subsurface structures, such as geothermal reservoir, cap rock, heat source (Figure 6). In surface and for the first hundred meters, geological map and previous geological observations have been integrated in the model. A previous electrical profile has been reinterpreted under the light of the new geological observation and integrated. This brings strong arguments for the two faults crosscutting the island: one by at the Airport Beach and another by Moya, where little circulation has been previous observed (Traineau *et al.*, 2006) (Figure 6).

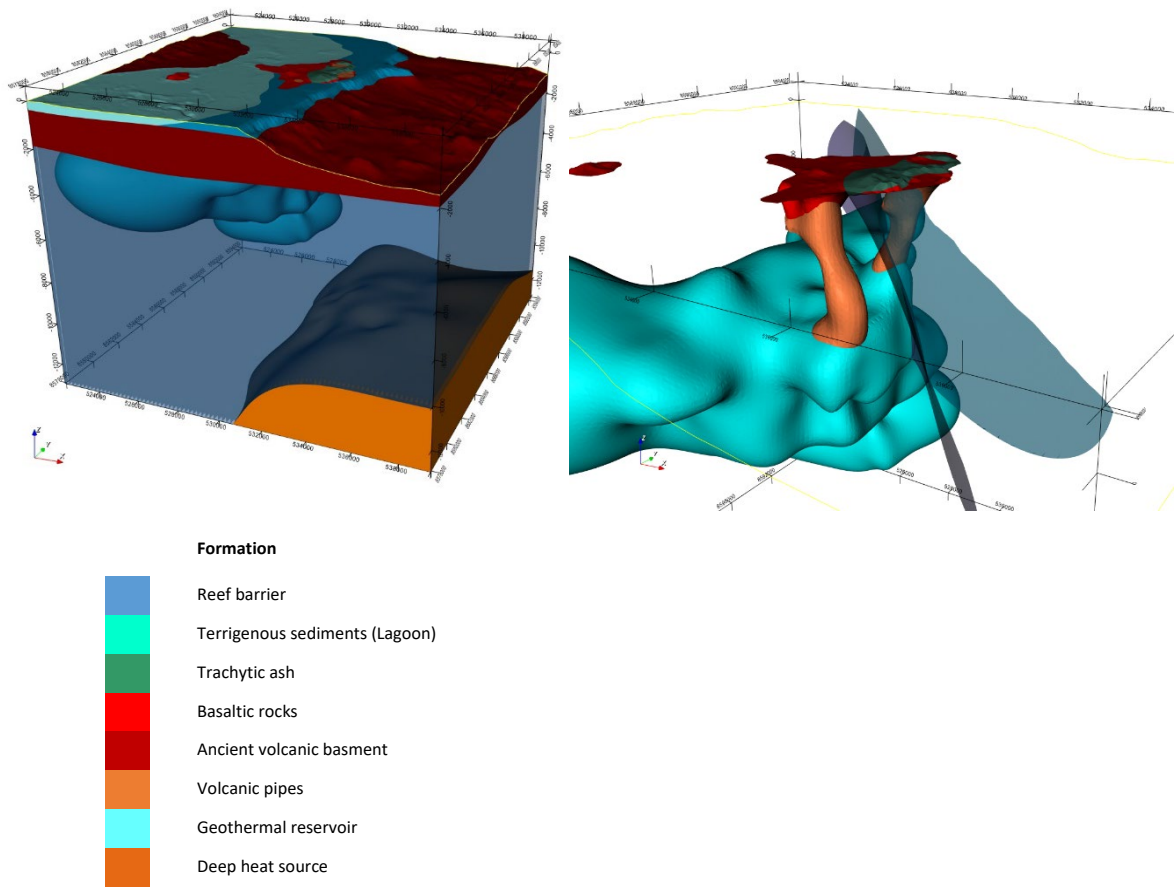

**Formation**

- Reef barrier
- Terrigenous sediments (Lagoon)
- Trachytic ash
- Basaltic rocks
- Ancient volcanic basment
- Volcanic pipes
- Geothermal reservoir
- Deep heat source

**Figure 6: 3D geological model of Petite Terre Island and its surrounding. Total depth of model is 12km. Y=North. Left: global view. Right: geothermal reservoir, pipes, and faults.**

Based on this geological model, the conceptual model define the hydrothermal behavior. The deep high conductivity zone constitutes the heat source. The resistive body constitutes the geothermal reservoir with a higher permeability than the basaltic rocks surrounding, with the upper high conductivity zone as the impermeable cap rock. Faults and pipes could drive deep geothermal fluid near the surface and have also been considered with a higher permeability. This conceptual model is using for hydrothermal simulations in order to define the best target to future drilling.

## 5. Conclusions

The geothermal exploration of Mayotte beginning in 2000's is still ongoing with new investigations and modelling of Petite Terre Island. This on-going work highlighted a geothermal reservoir at around 2 km depth with a temperature about 250°C, and a major NE-SW trend not mentioned previously. This trend constitute by faults and fractures could drive geothermal fluid near the surface.

A 3D geological and hydrothermal conceptual model has been performed based on whole data available. This is the base for a hydrothermal model and circulation simulations of deep geothermal fluid to near surface, which is still an ongoing work.

These hydrothermal simulations could give help to locate the best place for future exploration geothermal drilling.

## REFERENCES

Famin, V., Michon, L., Bourhane, A., 2020. The Comoros archipelago: a right-lateral transform boundary between the Somalia and Lwandle plates. Tectonophysics 789. https://doi.org/10.1016/j.tecto.2020.228539

Giggenbach, W.F., Sano, Y., Schmincke, H.U., 1991. CO2-rich gases from Lakes Nyos and Monoun, Cameroon; Laacher See, Germany; Dieng, Indonesia, and Mt. Gambier, Australia—variations on a common theme. J.Volcanol.Geotherm.Res. 45, 311–323. https://doi.org/10.1016/0377-0273(91)90065-8

S. Hautot, P. Tarits, K. Whaler, B. Le Gall, J.J. Tiercelin, C. Le Turdu. Deep structure of the baringo rift basin (Central Kenya) from three-dimensional magnetotelluric imaging: Implications for rift evolution. Journal of Geophysical Research: Solid Earth, 105 (2000), pp. 23493-23518

S. Hautot, R. Single, J. Watson, N. Harrop, D. Jerram, P. Tarits, K. Whaler, D. Dawes. 3-d magnetotelluric inversion and model validation with gravity data for the investigation of flood basalts and associated volcanic rifted margins. Geophys. J. Int., 170 (2007), pp. 1418-1430

Johnston J. M., Pellerin L., Hohmann G.W. (1992) - Evaluation of electromagnetic methods for geothermal reservoir detection. Geothermal resources transactions, vol. 16, oct. 92.

Lacquement, F., Nehlig, P., Bernard, 2013. Carte géologique de Mayotte. Feuille 1179. Edition BRGM.

Liuzzo, M., Di Muro, A., Rizzo, A.L., Caracausi, A., Grassa, F., Fournier, N., Shafik, B., Boudoire, G., Coltorti, M., Moreira, M., Italiano, F., 2021. Gas Geochemistry at Grande Comore and Mayotte Volcanic Islands (Comoros Archipelago), Indian Ocean. Geochemistry, Geophysics, Geosystems 22, e2021GC009870. https://doi.org/10.1029/2021GC009870.

Nehlig, P., Lacquement, F., Bernard, J., Audru, J.C., Caroff, M., Deparis, J., Jaouen, T., Pelleter, A.-A., Perrin, J., Prognon, C., Vittecoq, B., 2013. Notice de la carte géologique de Mayotte à 1/30 000.  Rapport BRGM RP-61803-FR. 135 p.

Pajot G.,Debeglia N. et Miehé J.-M. (2007) – Estimation du potentiel géothermique de Mayotte : Phase 2- Etape 1. Investigations géophysiques par gravimétrie, magnétisme et panneau de résistivité électrique. Rapport BRGM/RP-56027-FR, 59 p., 27 fig., 2 ann.

Sanjuan B., Baltassat J.-M., Bezelgues S., Brach M., Girard J.-F., Mathieu F., avec la collaboration de Debeglia N., Dupont F., François B., Miehé J.-M., Pajot G., Traineau H., 2008. Estimation du potentiel géothermique de Mayotte: Phase 2 - Etape 2. Investigation géologiques, géochimiques et géophysiques complémentaires, synthèse des résultats. Rapport BRGM/RP-56802-FR, 82 p.

Traineau H., Sanjuan B., Brach M., Audru J.-C., 2006. Etat des connaissances du potentiel géothermique de Mayotte. Rapport BRGM/RP-54700-FR, 81 p.